\documentclass[runningheads]{llncs}
\usepackage[T1]{fontenc}
\usepackage{wrapfig}
\usepackage{graphicx}
\usepackage{overpic}
\usepackage{graphicx}
\usepackage{color, colortbl}
\usepackage{booktabs}
\usepackage{overpic}
\usepackage{amsmath}
\usepackage{amsfonts}
\usepackage{amssymb}
\usepackage{hyperref}
\usepackage{caption}
\usepackage{listings}
\usepackage{graphicx}
\usepackage{wrapfig}
\usepackage{overpic}
\usepackage{lipsum}
\usepackage{mwe}
\usepackage{color}
\usepackage{multirow}
\usepackage{xcolor}
\usepackage{placeins} 
\usepackage{arydshln} 

\definecolor{bpcolor}{HTML}{D3E6F5}

\allowdisplaybreaks

\definecolor{pastelpurple}{RGB}{186, 148, 209}
\begin{document}
\title{Controlling Gender Bias in Retrieval via a Backpack Architecture}

\author{%
Amirabbas Afzali\thanks{Equal contribution}\inst{1} \and
Amirreza Velae$^\star$\inst{1} \and
Iman Ahmadi\inst{1}  \\ \and
Mohammad Aliannejadi\inst{2}
}

\authorrunning{Afzali et al.}

\institute{%
Sharif University of Technology\\ 
\email{\{amirabbas.afzali, amirreza.velae, iman.ahmadi\}@ee.sharif.edu}
\and
University of Amsterdam\\
\email{m.aliannejadi@uva.nl}
}

\maketitle              
\begin{abstract}
  The presence of social biases in large language models (LLMs) has become a significant concern in AI research. These biases, often embedded in training data, can perpetuate harmful stereotypes and distort decision-making processes. When LLMs are integrated into ranking systems, they can propagate these biases, leading to unfair outcomes in critical applications such as search engines and recommendation systems. Backpack Language Models, unlike traditional transformer-based models that treat text sequences as monolithic structures, generate outputs as weighted combinations of non-contextual, learned word aspects, also known as senses. Leveraging this architecture, we propose a framework for debiasing ranking tasks. Our experimental results show that this framework effectively mitigates gender bias in text retrieval and ranking with minimal degradation in performance. 

\end{abstract}

\vspace{-4mm}

\section{Introduction}

\vspace{-1mm}

Ranking and retrieval systems are essential components of modern information systems, enabling users to access relevant information efficiently. With the emergence of large language models (LLMs), the performance of these systems has significantly improved \cite{ma2023finetuningllamamultistagetext,zhuang2022rankt5finetuningt5text}. However, integrating LLMs into ranking pipelines introduces challenges related to fairness and bias. LLMs are trained on large internet-sourced datasets, which often contain social biases like gender \cite{nadeem-etal-2021-stereoset,kotek2023gender}. These biases can influence the model's decisions, leading to unfair outcomes. For example, a biased model may associate certain professions with specific genders, resulting in skewed rankings in hiring or recommendation scenarios.

Decades of research in psychology and sociology show that gender stereotypes systematically shape expectations and judgments, producing biased treatment and outcomes \cite{WhoWomenAre,Ellemers2018,article_Heilman,article_Swim}. In Information Retrieval (IR), neural representations can encode and amplify these stereotypes, leading to gender-skewed rankings; prior work analyzes these effects and introduces metrics to quantify gendered responses in ranked lists \cite{Rekabsaz_2020,sun2019mitigatinggenderbiasnatural,NIPS2016_a486cd07,fabris2020genderstereotypereinforcementmeasuring,fabris2020genderstereotypereinforcementmeasuring}. Large language models (LLMs) trained on web-scale corpora inherit similar societal biases \cite{nadeem-etal-2021-stereoset,kotek2023gender}, for example associating ``nurse'' with women and ``doctor'' with men \cite{zhao2018gender}, thereby reinforcing stereotypes \cite{navigli2023biases,omiye2023large}. When integrated into retrieval pipelines, these biases can be propagated and even amplified, posing risks in high-stakes settings such as hiring and healthcare \cite{10.1007/978-3-030-72240-1_18,otterbacher2018addressing,fairnessranking1,fairnessranking2}. 
Several out-of-process approaches have been proposed to mitigate gender bias, typically via post-processing methods such as re-ranking and embedding adjustments \cite{Zehlike_2017,10.1145/3299869.3300079,10.1145/3477495.3531891}.

\begin{wrapfigure}[21]{r}{0.5\columnwidth} 
  \vspace{-12mm}
  \centering
  \includegraphics[width=1.1\linewidth]{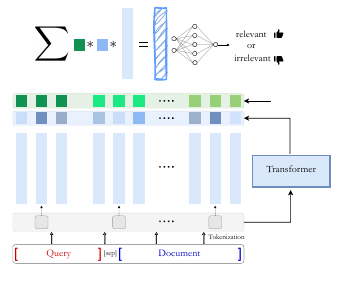}
    \put(-74,31){\scalebox{0.65}{$t_n$}}
    \put(-127.1,31){\scalebox{0.65}{$t_2$}}
    \put(-170,31){\scalebox{0.65}{$t_1$}}
  \put(-38,98){\scriptsize $\mathcal{M}_{\mathbf{s}}(\ell)$}
  \vspace{-4mm}
  \caption{Overview of our bias-controllable ranking framework. Tokens are decomposed into non-contextual sense vectors, whose sensitivities to a target aspect are measured. A policy \( \mathcal{M}_{\mathbf{s}}(\ell) \) reweights these senses, and the aggregated representations are passed through an MLP to yield the final relevance score.}
  \label{fig:bp}
  \vspace{7cm}
\end{wrapfigure}
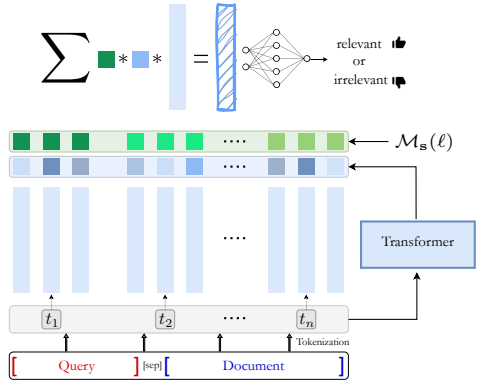

 \noindent These methods operate externally to the model and often require additional fine-tuning. However, building a framework that enables direct control over social bias within a ranking system, without retraining, remains an open challenge. This raises a critical question: \textbf{\textit{Can we integrate bias mitigation directly into the model's inference-time scoring pipeline to enable adaptive fairness control?}}\\

Backpack Language Models~\cite{hewitt2023backpacklanguagemodels} provide an interpretable architecture by representing each token as a weighted combination of non-contextual sense vectors. In this work, we propose a bias-aware ranking framework that leverages this structure to achieve rankings resilient to gender bias. Our method reweights the sense vectors based on fairness criteria directly within the model's inference pipeline, without requiring additional fine-tuning to ensure fairness. On MS~MARCO, our model pairs strong effectiveness with improved fairness: with $\lambda=1.0$ it achieves the best MRR@10/NDCG@10 ($0.3343/0.4025$), and with a modest fairness weight ($\lambda=0.5$) it attains the lowest RaB/ARaB across TF/Boolean at all cutoffs while staying within $\approx$2--3\% of peak effectiveness (Tables~\ref{tab:ranking_variants} and \ref{tab:gender_bias}).

\vspace{-2mm}

\section{Preliminaries}\label{sec:prelem}

\vspace{-2mm}

\vspace{1mm}
\noindent \textbf{Ranking Systems.}\; Given a query \(q_i\) and candidates \(\mathcal{D}_i=\{d_{i1},\ldots,d_{im}\}\), a ranker outputs a permutation of \(\mathcal{D}_i\). Supervision is a relevance vector \(y_i=(y_{i1},\ldots,y_{im})\) with \(y_{ij}\ge 0\) (binary or graded). We learn a parametric scorer \(f_{\theta}\) that assigns each pair \((q_i,d_{ij})\) a score \(\hat{y}_{ij}=f_{\theta}(q_i,d_{ij})\); sorting \(\{\hat{y}_{i1},\ldots,\hat{y}_{im}\}\) in descending order yields the ranking. Training encourages higher scores for documents with larger relevance.

\vspace{1mm}
\noindent \textbf{Backpack Architecture.}\; Let \(\mathcal{V}\) be a finite vocabulary.  The Backpack architecture operates as a mapping function that transforms an input token sequence \( x_{1:n} = (x_1, \ldots, x_n) \), where \( x_i \in \mathcal{V} \), into an output vector sequence \( o_{1:n} = (o_1, \ldots, o_n) \) with \( o_i \in \mathbb{R}^d \). For each \( x \in \mathcal{V} \), the Backpack model constructs \( k \) sense vectors \(C(x)_{1}, \ldots, C(x)_{k}\), where \( C : \mathcal{V} \to \mathbb{R}^{d \times k} \) defines the sense embedding space. These sense vectors are a multi-vector extension of traditional non-contextual word representations, capturing different aspects or senses of a word independently of context. This transformation is achieved through a weighted aggregation of \( k \) distinct sense vectors associated with each token. Formally, each output vector \( o_i \) is computed as:
\begin{equation}\label{eq:output}
  \footnotesize
  o_i = \sum_{j=1}^{n} \sum_{\ell=1}^{k} \alpha_{\ell ij} C(x_j)_\ell~.
\end{equation}
\noindent Here, $ \alpha \in \mathbb{R}^{k \times n \times n} $ represents contextualization weights generated by a function \( A: \mathcal{V}^n \to \mathbb{R}^{k \times n \times n} \) conditioned on the input sequence, i.e., \( \alpha = A(x_{1:n}) \). The symbol \( C(x_j)_\ell \in \mathbb{R}^d \) denotes the \( \ell^{\text{th}} \) sense vector for token \( x_j \).

The probabilistic framework of the Backpack model computes probabilities over some output space \( \mathcal{Y} \) through a log-linear transformation of the sequence representation \( o_{1:n} \in \mathbb{R}^{d \times n} \):
\begin{equation}
  \small
  p(y \mid o_{1:n}) = \mathrm{Softmax}\big(E( o_{1:n})\big)~,
\end{equation}
\noindent where \( E: \mathbb{R}^{d \times n} \to \mathbb{R}^{|\mathcal{Y}|} \) is a linear transformation and \( y \in \mathcal{Y} \). This formulation ensures that the output probabilities maintain a log-linear dependence on the sense vectors \( C(x_j)_\ell \), enabling explicit attribution of predictive influence to individual sense components across different contexts.

\vspace{-2mm}

\section{Methodology}
\vspace{-1mm}

In this section, we detail our proposed framework. To leverage the pretrained knowledge of the Backpack-LM for ranking tasks, we replace the final linear layer of the language model with a scalar output layer with sigmoid activation. We then fine-tune the pretrained model on the ranking task using the \textit{listwise softmax cross-entropy} loss \cite{10.1145/3341981.3344221}, which optimizes ranking by considering the entire list rather than treating documents individually or in pairs. For a given query \( q_i \), where \( y_{ij} \) represents the ground-truth relevance of document \( d_{ij} \) and \( \hat{y}_{ij} \) denotes our model's predicted score, the loss is defined as:
\begin{equation}
  \small
  \mathcal{L}_{\text{Softmax}}(y_i, \hat{y}_i) = - \sum_{j=1}^{m} y_{ij} \log \left( \frac{\exp(\hat{y}_{ij})}{\sum_{j'} \exp(\hat{y}_{ij'})} \right)~.
\end{equation}
During the inference phase, we aim to disentangle and control different aspects of language, including potential biases like gender bias. To this end, we propose a two-step procedure as follows:

\noindent \textbf{(i) Sense-Attribute Alignment.} 
Inspired by earlier work on Backpack Language Models~\cite{hewitt2023backpacklanguagemodels}, we use cosine similarity to measure how each sense $\ell$ aligns with a targeted social attribute (e.g., gender):
\begin{equation}
\footnotesize
  \mathrm{Sim}_\ell(x,x')=\mathrm{cossim}\!\big(C(x)_\ell,\; C(x')_\ell\big)~.
\end{equation}
Given an auxiliary set of polarity pairs $\mathcal{D}^{\mathrm{aux}}=\{(d_i^{-},d_i^{+})\}_i$ (e.g., \textit{he/she, man/woman},), we assign each sense $\ell\in\{1,\dots,k\}$ an attribute score:
\begin{equation}
\footnotesize
  s_\ell \;=\; \frac{1}{|\mathcal{D}^{\mathrm{aux}}|}\;
  \sum_{(d^{-},d^{+})\in \mathcal{D}^{\mathrm{aux}}}
  \mathrm{Sim}_\ell(d^{-},d^{+})~.
  \label{eq:sense-attr-score}
\end{equation}
We use cross-lingual gendered word pairs proposed by Nangia et al.~\cite{nangia2020crows}, as our auxiliary set. 
Lower (i.e., more negative) values of \( s_\ell \) indicate \textit{higher sensitivity} of sense \(\ell\) to the targeted attribute. This is because strongly negative cosine similarities between opposite-polarity pairs (e.g., \textit{he/she}, \textit{man/woman}) imply that their corresponding sense vectors point in nearly opposite directions. In other words, the more negative the average similarity \( s_\ell \) is, the more clearly sense \(\ell\) separates the two poles of the attribute, indicating stronger sensitivity to that aspect. \looseness=-1

\vspace{1mm}
\noindent \textbf{(ii) Sense-Level Reweighting for Bias Control.} 
We define \(\mathcal{M}_{\mathbf{s}} : \{1, \dots, k\} \to \mathbb{R}^{+}\) as a mapping function that assigns a positive weight to each sense index, based on the set attribute scores computed previously. When \(\mathcal{M}_{\mathbf{s}}(\ell) < 1\), the influence of the \(\ell\)-th sense is suppressed; otherwise, it is amplified. Finally, we compute the debiased output vector \(\Tilde{o}_i\) as a weighted variant of the original Backpack output vector (Eq.~\ref{eq:output}):
\begin{equation}
  \vspace{-3mm}
  \small
  \Tilde{o}_i = \sum_{j=1}^{n} \sum_{\ell=1}^{k} \mathcal{M}_{\mathbf{s}}(\ell) \, \alpha_{\ell ij} \, C(x_j)_\ell~.
  \vspace{-1mm}
\end{equation} 
By adjusting the mapping function \(\mathcal{M}_{\mathbf{s}}\), we can directly control the contribution of the targeted social attribute (e.g., gender) in the final representation. We then pass the reweighted representation \(\Tilde{o}_i\) through a two-layer MLP to produce a scalar score representing the document's predicted relevance. Selectively adjusting specific sense vectors via \(\mathcal{M}_{\mathbf{s}}\) can steer the scoring function toward greater gender neutrality or higher task accuracy, depending on the desired trade-off. In our experiments, the mapping function outputs a weight of \(1\) for all senses except the two most gender-sensitive ones, for which it outputs a value \(\lambda < 1\). These two senses are identified based on their gender attribute scores, and their influence is scaled by \(\lambda\), with smaller values corresponding to stronger suppression of gender-related information.

\begin{figure*}[t]
  \centering
    \begin{minipage}{0.45\textwidth}
      \centering
      \captionof{table}{%
      Ranking performance on the MS MARCO passage dev set. Decreasing the value of \( \lambda \) slightly reduces ranking performance, highlighting the fairness–performance trade-off.
      }
      \vspace{-2mm}
      \resizebox{\linewidth}{!}{%
        \begin{tabular}{l|c|c}
          \toprule
          Backbone & MRR@10 & NDCG@10 \\
          \midrule
          BM25        & 0.194 & 0.241 \\
          PACRR & 0.259& 0.313 \\
          MP          & 0.249 & 0.301 \\
          KNRM        & 0.235 & 0.288 \\
          ConvKNRM    & 0.277 & 0.332 \\
          \midrule
          GPT-2       & 0.316 & 0.384 \\
          \rowcolor{bpcolor} Ours ($\lambda = 1.0$) & \textbf{0.334} & \textbf{0.402} \\
          \hdashline
          \rowcolor{bpcolor} Ours ($\lambda = 0.5$) & 0.325 & 0.395 \\
          \rowcolor{bpcolor} Ours ($\lambda = 0.7$) & 0.330     & 0.399 \\
          \bottomrule
        \end{tabular}
      }
      \label{tab:ranking_variants}
    \end{minipage}
  \hfill
  \begin{minipage}{0.52\textwidth}
    \centering
    \includegraphics[width=\linewidth]{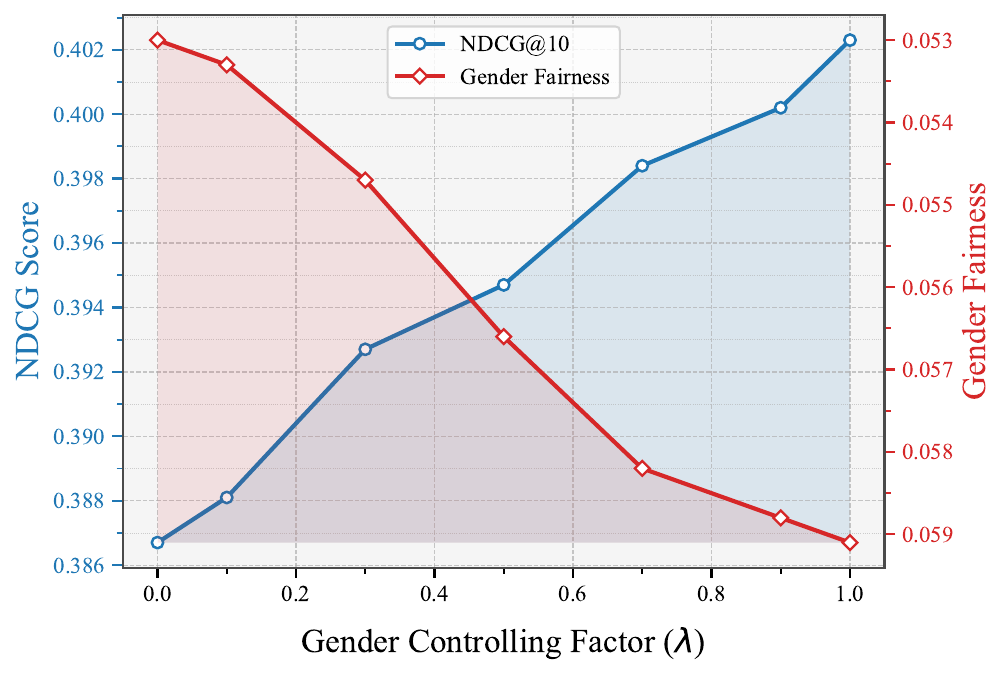}
    \vspace{-2mm}
    \caption{%
      Effect of the control weight \( \lambda \) on the trade-off between ranking effectiveness (NDCG@10) and gender fairness, quantified as \(1 - \mathrm{ARaB}\).
    }
    \label{fig:effect_alpha}
  \end{minipage}
  \vspace{-4mm}
\end{figure*}

\vspace{-2mm}

\section{Experiments}
\label{sec:exp}
\vspace{-2mm}

\noindent \textbf{Experimental Setup.} Our experiments focus on controlling gender bias in document ranking. \textbf{\textit{Model architecture:}} We begin by evaluating the general ranking performance of our proposed architecture, which integrates a 170M-parameter pretrained Backpack-LM\footnote{\href{https://huggingface.co/stanfordnlp/backpack-gpt2}{https://huggingface.co/stanfordnlp/backpack-gpt2}} with a scalar output layer with sigmoid activation (see Fig.~\ref{fig:bp}). \textbf{\textit{Dataset:}} All models are fine-tuned on the training split of the MS~MARCO Passage Ranking dataset~\cite{bajaj2018msmarcohumangenerated}, which includes approximately 530{,}000 training queries and 6{,}800 development queries, and each query is annotated with binary relevance labels. For evaluation of general ranking ability (Table~\ref{tab:ranking_variants}), we use the official development set and the top-100 BM25-retrieved passages obtained from Pyserini's prebuilt index~\cite{lin2021pyserinieasytousepythontoolkit}. Training is performed for 4 epochs with a learning rate of \( 1 \times 10^{-5} \). \textbf{\textit{Evaluation:}} We evaluate general ranking using commonly used metrics such as NDCG@10 and MRR@10, which accounts for graded relevance and position-based discounting. To assess gender bias in the ranking outputs, we follow the evaluation setting introduced by~\cite{Rekabsaz_2020}. Following their approach, we filter out queries with explicit gender terms, resulting in 1{,}765 non-gendered queries. We compare our method against widely used lexical and neural ranking baselines: \textit{BM25}~\cite{robertson2009probabilistic} by its default parameters $k_1{=}0.9$, $b{=}0.4$. \textit{KNRM}~\cite{Chenyan2017EndtoEndNR}, \textit{MP}~\cite{Pang_Lan_Guo_Xu_Wan_Cheng_2016}, \textit{PACRR}~\cite{hui-etal-2017-pacrr}, \textit{ConvKNRM}~\cite{dai2018convolutional}: all use 300-dimensional GloVe embeddings~\cite{pennington2014glove} and follow hyperparameters from their original codebases. The framework consists of two elements: first, defining document gender magnitude; second, quantifying gender bias in rankings. Let \( G_f{=}\{\textit{she}, \textit{woman}, \textit{her}\} \) and \( G_m{=}\{\textit{he}, \textit{man}, \textit{him}\} \) denote female- and male-associated terms. For a document \( d \), the female and male gender magnitudes, \( \text{mag}_f(d) \) and \( \text{mag}_m(d) \), are computed via two variants:

\vspace{-2mm}

\FloatBarrier
\begin{table}[t]
  \centering
  \renewcommand{\arraystretch}{1.15}
  \setlength{\tabcolsep}{7pt}
  \caption{Evaluation of gender bias in retrieval rankings measured by Rank Bias (RaB) and Average Rank Bias (ARaB). Lower values indicate less bias. Our model with \(\alpha=0.5\) consistently achieves the lowest bias across all metrics and all cut-offs. Highlighted models are our proposed method.}
  \label{tab:gender_bias}
  \vspace{1mm}
  \scalebox{0.72}{%
    \begin{tabular}{lcccc||lcccc}
      \toprule
                                                 & \multicolumn{2}{c}{\textbf{TF}}          & \multicolumn{2}{c}{\textbf{Boolean}}
                                                 &                                          & \multicolumn{2}{c}{\textbf{TF}}      & \multicolumn{2}{c}{\textbf{Boolean}}                                                                                                            \\
      \cmidrule(lr){2-3} \cmidrule(lr){4-5} \cmidrule(lr){7-8} \cmidrule(lr){9-10}
      \textbf{Model}                             & RaB                                      & ARaB                                 & RaB                                  & ARaB
                                                 & \textbf{Model}                           & RaB                                  & ARaB                                 & RaB            & ARaB                                                                                    \\
      \midrule
      &\multicolumn{4}{c||}{\textbf{Cut-off: 10}} && \multicolumn{4}{c}{\textbf{Cut-off: 30}}                                                                                                                                                                                          \\
      \midrule
      BM25                                       & 0.062                                    & 0.063                                & 0.048                                & 0.044
                                                 & BM25                                     & 0.058                                & 0.060                                & 0.048          & 0.047                                                                                   \\
      PACRR                                      & 0.080                                    & 0.084                                & 0.062                                & 0.063
                                                 & PACRR                                    & 0.070                                & 0.078                                & 0.057          & 0.060                                                                                   \\
      MP                                         & 0.065                                    & 0.072                                & 0.052                                & 0.056
                                                 & MP                                       & 0.059                                & 0.066                                & 0.049          & 0.053                                                                                   \\
      KNRM                                       & 0.067                                    & 0.064                                & 0.051                                & 0.051
                                                 & KNRM                                     & 0.068                                & 0.067                                & 0.055          & 0.053                                                                                   \\
      ConvKNRM                                   & 0.080                                    & 0.077                                & 0.064                                & 0.060
                                                 & ConvKNRM                                 & 0.069                                & 0.074                                & 0.057          & 0.059                                                                                   \\
      \rowcolor{bpcolor}
      Ours ($\alpha=1$)                          & 0.064                                    & 0.064                                & 0.047                                & 0.047          & Ours ($\alpha=1$)   & 0.061          & 0.061          & 0.048          & 0.046          \\
      \rowcolor{bpcolor}
      Ours ($\alpha=0.5$)                        & \textbf{0.053}                           & \textbf{0.056}                       & \textbf{0.039}                       & \textbf{0.042} & Ours ($\alpha=0.5$) & \textbf{0.052} & \textbf{0.053} & \textbf{0.041} & \textbf{0.041} \\
      \midrule
      & \multicolumn{4}{c||}{\textbf{Cut-off: 20}} && \multicolumn{4}{c}{\textbf{Cut-off: 40}}                                                                                                                                                                                          \\
      \midrule
      BM25                                       & 0.060                                    & 0.062                                & 0.048                                & 0.046
                                                 & BM25                                     & 0.057                                & 0.060                                & 0.048          & 0.047                                                                                   \\
      PACRR                                      & 0.073                                    & 0.081                                & 0.058                                & 0.061
                                                 & PACRR                                    & 0.066                                & 0.076                                & 0.055          & 0.059                                                                                   \\
      MP                                         & 0.063                                    & 0.068                                & 0.052                                & 0.054
                                                 & MP                                       & 0.055                                & 0.064                                & 0.045          & 0.051                                                                                   \\
      KNRM                                       & 0.068                                    & 0.066                                & 0.054                                & 0.052
                                                 & KNRM                                     & 0.067                                & 0.067                                & 0.056          & 0.054                                                                                   \\
      ConvKNRM                                   & 0.071                                    & 0.075                                & 0.058                                & 0.059
                                                 & ConvKNRM                                 & 0.068                                & 0.073                                & 0.056          & 0.059                                                                                   \\
      \rowcolor{bpcolor}
      Ours ($\alpha=1$)                          & 0.058                                    & 0.062                                & 0.045                                & 0.046          & Ours ($\alpha=1$)   & 0.059          & 0.061          & 0.048          & 0.047          \\
      \rowcolor{bpcolor}
      Ours ($\alpha=0.5$)                        & \textbf{0.051}                           & \textbf{0.053}                       & \textbf{0.040}                       & \textbf{0.041} & Ours ($\alpha=0.5$) & \textbf{0.053} & \textbf{0.053} & \textbf{0.043} & \textbf{0.041} \\
      \bottomrule
    \end{tabular}
  }
\end{table}

\[
\begin{aligned}
\footnotesize
\text{mag}_f^{\text{TF}}(d) &= \sum_{\substack{w \in G_f \\ \#(w,d) > 0}} \log\!\big(\#(w,d)\big)~,
\qquad
\text{mag}_f^{\text{Bool}}(d) =
\begin{cases}
1, & \text{if } \sum_{w \in G_f} \#(w,d) > 0~,\\
0, & \text{otherwise~.}
\end{cases}
\end{aligned}
\]

\noindent For a query \( q \) with ranked list \(\{d(q)_i\}_{i=1}^t\), gender bias is measured using RaB and ARaB:
\[
\begin{aligned}
 \footnotesize \text{RaB}_t(q) = \frac{1}{t} \sum_{i=1}^{t} \big( \text{mag}_f(d(q)_i) - \text{mag}_m(d(q)_i) \big)~, 
\quad
 \footnotesize \text{ARaB}_t(q) = \frac{1}{t} \sum_{x=1}^{t} \text{RaB}_x(q)~.
\end{aligned}
\]
\noindent We report \(\text{RaB}_t\) and \(\text{ARaB}_t\) as averages over all queries.
\vspace{1.5mm}

\noindent \textbf{Results.}\; For a controlled comparison, we train a size-matched GPT-2 ranker and our Backpack-based ranker under identical training protocols. We evaluate three values of the controlling parameter \(\lambda\): \(\lambda{=}1.0\) (no mitigation), \(\lambda{=}0.7\) (moderate mitigation), and \(\lambda{=}0.5\) (strong mitigation). As shown in Table~\ref{tab:ranking_variants}, our method with \(\lambda=1.0\) achieves the best effectiveness, outperforming the size-matched GPT-2 by \(+5.7\%\) MRR and \(+4.7\%\) NDCG, and surpassing standard lexical (BM25) and neural baselines (e.g., KNRM, PACRR, ConvKNRM) by a substantial margin. Reducing \(\lambda\) leads to a slight drop in ranking quality, yet both \(\lambda=0.7\) and \(\lambda=0.5\) settings remain clearly superior to all classical baselines, illustrating a controllable fairness–performance trade-off.

Table~\ref{tab:gender_bias} reports RaB and ARaB metrics at different cut-off values. The results highlight the effectiveness of our method in mitigating gender bias. In particular, our approach with \(\lambda = 0.5\) consistently outperforms all baselines across both metrics. Note that when \(\lambda = 1.0\), the mapping function reduces to the identity, and no sense-level reweighting is applied, meaning that their associated biases remain unchanged. Fig.~\ref{fig:effect_alpha} illustrates the trade-off between gender fairness and ranking performance as the value of \( \lambda \) is varied. As expected, decreasing \( \lambda \) improves fairness by reducing bias, with only a marginal decrease in ranking quality. When \( \lambda = 1 \), the mapping function is the identity and no sense-level reweighting is applied (i.e., \( \mathcal{M}_{\mathbf{s}}(\ell)=1 \) for all \( \ell \)), so the original sense contributions remain unchanged.  \looseness=-1

\vspace{-3mm}
\section{Conclusion}
\vspace{-3mm}
We propose a bias-controllable ranking framework based on Backpack Language Models that mitigates gender bias during inference via sense vector reweighting. Experiments show that it significantly reduces gender bias while maintaining ranking performance, offering a practical solution for fair information retrieval. In the future, we plan to test this architecture on larger models and test its generalization on other tasks and datasets.

\bibliographystyle{splncs04}
\bibliography{custom}

\end{document}